# A Note On Stability of Steady Motion of a Rolling Disk

Milan Batista

University of Ljubljana
*Faculty of Maritime Studies and Transportation*
*Pot pomorscakov 4, 6320 Portoroz, Slovenia*
milan.batista@fpp.edu

(July 15, 2005)

## Abstract

This note adds some critical remarks on the discussion presented in the McDonald's paper ([1]) on stability of steady motion of the well known problem of a disk rolling on a rough horizontal plane.

In McDonald's paper ([1]) the general analysis of small oscillations about steady motion is based on equation (equation 54 in [1])

$$\frac{\varpi^2}{\Omega_0^2}(k+1) = 3k\cos^2\alpha_0 + \sin^2\alpha_0 + \frac{b}{a}\left[(6k+1)\cos\alpha_0 - (2k+1)\frac{\sin^2\alpha_0}{\cos\alpha_0}\right] \\ + 2\frac{b^2}{a^2}(2k+1) \quad (1)$$

but it seems that there is some incorrectness in their analysis. To clarify this, equation (1) is rewritten in the form

$$\lambda^2 = Ax^2 + Bx + C \quad (2)$$

where

$$\lambda^2 \equiv \frac{\varpi^2}{\Omega_0^2}(k+1)$$
$$x \equiv \frac{b}{a}$$
$$A \equiv 2(2k+1) \quad (3)$$
$$B \equiv (6k+1)\cos\alpha_0 - (2k+1)\frac{\sin^2\alpha_0}{\cos\alpha_0}$$
$$C \equiv 3k\cos^2\alpha_0 + \sin^2\alpha_0$$

In the next step of the analysis the authors calculate the angle at which *B* becomes negative

$$B \leq 0 \quad \Rightarrow \quad \alpha_0 \geq \arctan\sqrt{\frac{6k+1}{2k+1}} \qquad (4)$$

For a uniform disk with $k = \frac{1}{4}$ this angle is $\alpha_0 \geq \arctan\frac{\sqrt{15}}{3} = 52.2^0$ (which differs from the value of $60^0$ reported by the authors). From this the authors conclude that »for positive $b$ the motion is unstable for large $\alpha_0$ and that the disk will appear fall over quickly into a rolling motion with $\alpha_0 \leq 60^0$«. It seems that this is slightly incorrect for the following reason. Namely, the rolling will be stable, if $\lambda^2 > 0$. For fixed $\alpha_0$ this means that (2) must have no real zeros which imply that its discriminant of (2) must satisfy the condition $B^2 - 4AC < 0$. The limit is at

$$B^2 - 4AC = 0 \quad \Rightarrow \quad \alpha_0^{(c)} = \pm\arctan\frac{\sqrt{(2k+1)\left(6k+5+2\sqrt{6(k+1)(2k+1)}\right)}}{2k+1} \qquad (5)$$

which has the value $\alpha_0^{(c)} = \arctan\frac{\sqrt{39+18\sqrt{5}}}{3} \simeq 1.24\left(\simeq 71.4^0\right)$ for a disk ($k = 1/2$) and $\alpha_0^{(c)} \simeq 70.8^0$ for a hoop ($k = 1/4$). The same value for $\alpha_0^{(c)}$ for a disk (obtained by a different procedure) is reported by Kuleshov ([2]).

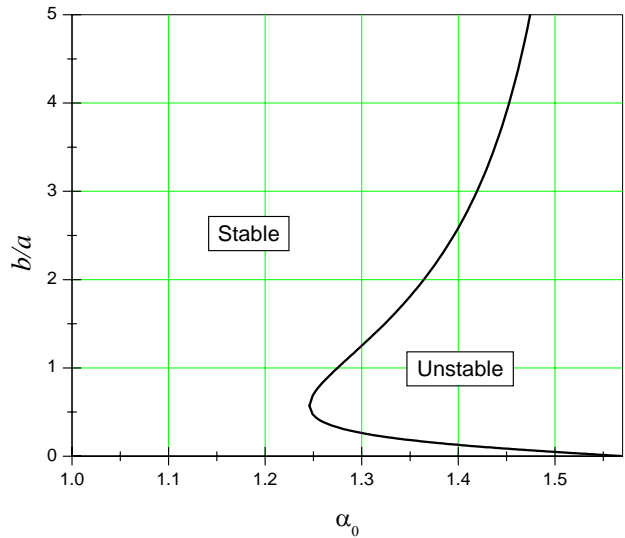

**Figure 1.** Stability regions for a disk.

The graph of $Ax^2 + Bx + C = 0$ and associated stability regions for a disk are shown on Figure 1. Note that the value $\alpha_0 = 1.24$ represents the turning point of the graph. Note also that for very small and as well very large values of $b$ the motion is also stable for large $\alpha_0$.